\documentclass[onecolumn,showpacs,preprintnumbers,amsmath,amssymb]{revtex4}
\usepackage{amsmath,amssymb,graphics,epsfig,subfigure}
\usepackage{color}
\usepackage[colorlinks,linkcolor=blue,anchorcolor=blue,citecolor=green]{hyperref}

\begin{document}

\title{Weak cosmic censorship conjecture for Kerr-Taub-NUT black hole with test scalar field and particle}

\author{Si-Jiang Yang$^a$, %\footnote{yangsj18@lzu.edu.cn
        Jing Chen$^a$, %\footnote{chenj19@lzu.edu.cn},
        Jun-Jie Wan$^a$, %\footnote{wanjj19@lzu.edu.cn},
        Shao-Wen Wei$^a$, %\footnote{weishw@lzu.edu.cn},
        Yu-Xiao Liu$^a$$^b$\footnote{liuyx@lzu.edu.cn, corresponding author}}

\affiliation{$^{a}$Institute of Theoretical Physics $\&$ Research Center of Gravitation, Lanzhou University, Lanzhou 730000, China\\
$^{b}$Key Laboratory for Magnetism and Magnetic of the Ministry of Education, Lanzhou University, Lanzhou 730000, China}

\date{\today}

\begin{abstract}
   Motivated by the recent researches of black holes with NUT charge, we investigate the validity of the weak cosmic censorship conjecture for Kerr-Taub-NUT black hole with a test massive scalar field and a test particle, respectively. For the scalar field scattering gedanken experiment, we consider an infinitesimal time interval process. The result shows that both extremal and near-extremal Kerr-Taub-NUT black holes cannot be over-spun. For the test particle thought experiment, the study suggests that extremal Kerr-Taub-NUT black hole cannot be over-spun; while near-extremal Kerr-Taub-NUT black hole can be over-spun. By comparing the two methods, the results indicate the time interval for particles crossing the black hole horizon might be important for consideration of the weak cosmic censorship conjecture.
\end{abstract}

%\keywords{Black holes}


\maketitle

\section{Introduction}

In order to preserve the predictability of classical gravity theory, Penrose proposed the weak cosmic censorship conjecture \cite{Penr69}, which states that space-time singularities arising in gravitational collapse should be always hidden behind black hole event horizons and can never be seen by distant observers. The conjecture has become one of the cornerstones of black hole physics. A general proof of the conjecture is still beyond reach and it is one of the most outstanding unresolved problems in classical gravity theory \cite{JoMa11}. It is known that there are some methods to test the conjecture, such as numerically analyzing gravitational collapse of a scalar field or other matters
%a dust cloud or perfect fluid
\cite{Chop93,Chri84,OrPi87,ShTe91,Lemo92,East19}, numerical simulations of collision and merger of two black holes \cite{SCPBHY09,AELL19a,AELL19b}, analytically and numerically evolving perturbed black rings or black holes \cite{HeHM04,LePr10,FiKT16,FKLT17,CrSa17,EpGS19}. One way we are interested in is to ask whether it is possible to destroy a black hole horizon.

A classical thought experiment to destroy a black hole horizon by throwing large angular momentum and charged test particles into an extremal Kerr-Newman black hole was first proposed by Wald. The investigation showed that particles which would cause the destruction of the black hole horizon will not be captured by the black hole. This suggests that the horizon of an extremal Kerr-Newman black hole cannot be destroyed by test particles \cite{Wald74}. The result is supported by similar but systematic works of Rocha and Cardoso et al. for BTZ black holes \cite{RoCa11}, higher-dimensional Myers-Perry family of rotating black holes and a large class of five-dimensional black rings \cite{LCNR10}. While, further investigations turn out that the test particle approximation actually allows a black hole to ``jump over'' the extremal limit \cite{Hod02}. Hubeny showed that a near-extremal charged black hole can be over-charged by capturing a charged test particle \cite{Hube99}. By extending Hubeny's analysis, Jacobson and Sotiriou found that, for a near-extremal Kerr black hole, particle capture can indeed over-spin the black hole and create a naked singularity in the absence of back-reaction effects \cite{JaSo09}. The result violates the cosmic censorship conjecture. However, once radiation and self-force are taken into account, it is inspiring that these effects can prevent the formation of naked singularities. Radiation reaction can affect some of the orbits, and self-force may make comparable effects to the terms giving rise to naked singularities \cite{BaCK10,BaCK11,ZVPH13,CoBa15,CBSM15}.

Another intriguing way to destroy a black hole horizon is the scattering of classical or quantum fields from a black hole. Such scattering provides unexpected features due to super-radiance. In the process of super-radiance, the fields extract energy from the charged or rotational black hole. Super-radiance could prohibit the dangerous wave modes to be absorbed by the black hole. Semiz showed that classical complex massive scalar fields cannot destroy extremal dyonic Kerr-Newman black holes \cite{Semi11}. Further exploration of Gwak indicates that both extremal and near-extremal Kerr-(anti) de Sitter black holes cannot be over-spun by classical fields \cite{Gwak18}. Following this line, a series of works have shown that BTZ black holes \cite{Chen18,ZeHC19}, four-dimensional and higher-dimensional charged AdS black holes \cite{ChZY19,Gwak19} cannot be over-spun or over-charged by classical test fields. These results suggest that the weak cosmic censorship conjecture is preserved in classical wave scattering process. However, quantum mechanically, near-extremal black holes may capture dangerous quanta to become naked singularities due to quantum tunneling process. By ignoring back-reaction effects on background space-time, Matsas and Silva showed that quantized Klein-Gordon field can over-spin a near-extremal charged black hole to become a Kerr-Newman naked singularity, thus the weak cosmic censorship conjecture is violated \cite{MaSi07}. When back-reaction effects are taken into account, the fields would trigger the black hole to rotate, and the Reissner-Nordstr\"om space-time would become a charged slowly rotating black hole background. Hod showed that a quantized scalar field cannot destroy the horizon of the Reissner-Nordstr\"om black hole \cite{Hod08}. But further investigations suggest that the weak cosmic censorship conjecture may indeed be violated by quantized wave scattering process \cite{RiSa08,MRSSV09,RiSa11}.

Recently, black holes with NUT charge have attracted some interesting researches, such as gravitational lensing \cite{WLFY12}, thermodynamics and phase transition \cite{HeKM19,BGHK19,BGHK19b,WuWu19,ChJi19,John14}, strong cosmic censorship conjecture \cite{RaMC20} and weak cosmic censorship conjecture \cite{Duzt18}. In this paper, we investigate the validity of the weak cosmic censorship conjecture for Kerr-Taub-NUT black holes by considering scattering of a test massive scalar field and injection of a test particle, respectively. For scattering of a scalar field, our result suggests that both extremal and near-extremal Kerr-Taub-NUT black holes cannot be destroyed. For test particle injection, particles leading to violation of the weak cosmic censorship conjecture can not be captured by the extremal black hole, but can be captured by the near-extremal black hole. Therefore, the particles would destroy the horizon of the near-extremal black hole.

The outline of the paper is as follows. In Sec.\,\ref{2}, we briefly review the Kerr-Taub-NUT black hole and its thermodynamics. In Sec.\,\ref{3}, we explore the scattering of a massive scalar field in Kerr-Taub-NUT black hole background and obtain the energy and angular momentum of the scalar field. In Sec.\,\ref{4}, we test the validity of the weak cosmic censorship conjecture for the extremal and near-extremal Kerr-Taub-NUT black holes by considering scattering of the scalar field. In Sec.\,\ref{5}, we check the validity of the weak cosmic censorship conjecture by investigating injection of a test particle. The last section is devoted to discussions and conclusions.

\section{Kerr-Taub-NUT black hole and its thermodynamics}\label{2}

Kerr-Taub-NUT black hole is a four-dimensional rotating black hole. It is a vacuum solution of the  Einstein's  equation.
In Boyer-Lindquist coordinates $(t, r, \theta, \phi)$, the Kerr-Taub-NUT metric can be written as
\begin{equation}
    ds^2=-\frac{\Delta}{\Sigma}\left[dt+(2n\cos \theta -a \sin^2\theta)d\phi \right]^2+\frac{\Sigma}{\Delta}dr^2+ \frac{\sin^2\theta}{\Sigma}\left[adt-(r^2+a^2+n^2)d\phi \right]^2+ \Sigma d\theta^2,\label{KTNmetric}
\end{equation}
with the metric functions
\begin{eqnarray}
  \Delta &=& r^2-2Mr+a^2-n^2, \\
  \Sigma &=& r^2+(n+a\cos \theta )^2.
\end{eqnarray}
where $M$ , $a$ and $n $ are the mass, rotation parameter and NUT parameter respectively.

The above metric describes a black hole where the Misner strings are symmetrically distributed on the north and south poles. Space-time singularity occurs for $\Sigma=0$. For $a^2<n^2$, $\Sigma $ is always positive. In this case, there is no space-time singularity and the black hole is regular \cite{Mill73}. For $a^2\geq n^2$, space-time singularity occurs. In this paper, we only consider the later case.
\begin{figure}
  \centering
  \includegraphics[width=8cm]{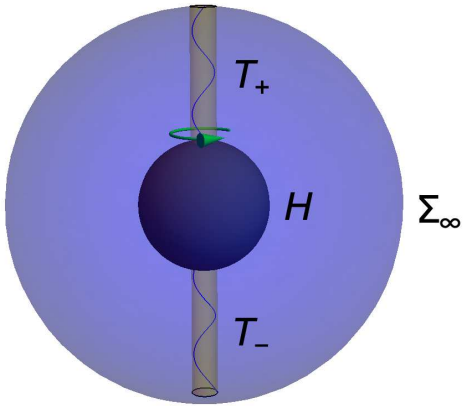}
  \caption{Kerr-Taub-NUT boundaries: Misner turbes \cite{BGHK19b}. Apart from the standard boundary  the event horizon $H $ and spatial infinity $\Sigma_\infty$ , Kerr-Taub-NUT space-time has two Misner tubes $T_{\pm} $  located at north and south pole.}\label{fig1}
\end{figure}

For a non-extremal Kerr-Taub-NUT black hole, two horizons occur on the surfaces
\begin{equation}\label{horizon}
    \Delta=r^2-2Mr+a^2-n^2=0,
\end{equation}
which gives the outer and inner horizons
\begin{equation}\label{horiz}
    r_{\pm}=M\pm \sqrt{M^2+n^2-a^2}.
\end{equation}
The outer horizon corresponds to the event horizon.
For an extremal black hole, the two horizons coincide and the degenerate horizon locates at $r_{\text{ex}}=M$. The horizon disappears for
  $M^2+n^2<a^2$ .  In this case, the metric describes a naked singularity.
In the following, we will denote the event horizon with $r_\text{h}^{~}$.
The Kerr-Taub-NUT black hole solution reduces to a Taub-NUT regular black hole when $a=0$, and the solution becomes  a Kerr black hole solution in the absence of the NUT charge.

%It is explicit that the metric (\ref{KTNmetric}) describes a four-dimensional black hole carrying three parameters, mass $M$, NUT parameter $n$ and angular parameter $a$.

The area of the event horizon of the Kerr-Taub-NUT black hole is
\begin{equation}\label{Area}
    A=4\pi (r_\text{h}^2+a^2+n^2),
\end{equation}
while its temperature reads
\begin{equation}\label{temp}
  T=\frac{r_\text{h}^{~}-M}{2\pi(r_\text{h}^2+a^2+n^2)}.
\end{equation}
The angular velocity  of the black hole is  given by
\begin{equation}
    \Omega_\text{h}=\frac{a}{r_\text{h}^2+a^2+n^2}. \label{angular}
\end{equation}

The thermodynamics of black holes with NUT parameter have been investigated
extensively \cite{HeKM19,BGHK19,BGHK19b,WuWu19,ChJi19}.
The first law of thermodynamics for the Kerr-Taub-NUT black hole is \cite{BGHK19}
\begin{equation}\label{firs}
  dM=TdS+\Omega_\text{h}dJ+\psi dN,
\end{equation}
where the angular momentum $J$ and the thermodynamical charge $N$ of the black hole are
\begin{eqnarray}
  J &=& (M+\frac{n^2}{r_\text{h}^{~}})a,  \label{J}\\
  N &=& -\frac{4\pi n^3}{r_\text{h}^{~}}, \label{N}
\end{eqnarray}
and $\psi$ is the Misner potential
\begin{equation}\label{Misner}
  \psi=\frac{1}{8\pi n}.
\end{equation}

\section{Massive scalar field in Kerr-Taub-NUT space-time}\label{3}

\subsection{The scattering for massive scalar field}

The dynamics of a minimally coupled massive scalar field $\Psi$ with mass $\mu$ in the Kerr-Taub-NUT space-time is governed by the Klein-Gordon equation
\begin{equation}\label{field}
   \frac{1}{\sqrt{-g}}\partial_\mu\left(\sqrt{-g}g^{\mu\nu}\partial_\nu \Psi\right)-\mu^2\Psi=0.
\end{equation}
To make the whole problem more tractable, it is convenient to make the following decomposition for the scalar field \cite{Teuk72,Teuk73}
\begin{equation}\label{wavefun}
   \Psi(t,r,\theta,\phi)=e^{-i\omega t} R_{lm}(r)S_{lm}(\theta)e^{im\phi},
\end{equation}
where $ S_{lm}(\theta) $ are spheroidal angular functions, and the azimuthal number $ m $ is an integer.
Inserting this into the Klein-Gordon equation, we get the angular part of the equation
\begin{equation}
   \frac{1}{\sin\theta}\frac{d}{d\theta}\left( \sin\theta\frac{dS_{lm}}{d\theta}\right)- \left[ \frac{(2n\omega\cos\theta-a\omega\sin^2\theta+m)^2}{\sin^2\theta}+ \mu^2(n+a\cos\theta)^2-\lambda_{lm} \right]S_{lm}=0,
\end{equation}
and the radial part
\begin{equation}\label{radial}
   \frac{d}{dr}\left(\Delta\frac{dR_{lm}}{dr}\right)+\left(\frac{G^2}{\Delta} -\mu^2r^2+ 2am\omega-\lambda_{lm}\right)R_{lm}=0,
\end{equation}
where $\lambda_{lm}$ is the separation constant and is given by $ \lambda_{lm}=l(l+1)+{\cal O}(a^2\omega^2)$ \cite{Seid89}, and
\begin{equation}
   G=\omega(r^2+n^2+a^2)-am.
\end{equation}
The solutions to the angular part of the equation are the spheroidal angular functions \cite{Seid89}. Here, we are more concerned with the radial part. The contribution of the angular part in Eq.\,(\ref{angular}) will be reduced to unity in the fluxes by normalization condition \cite{Gwak18}. The equation for the radial part can be simplified by introducing the tortoise coordinate with the definition
\begin{equation}\label{tortoise}
   \frac{dr}{dr_*}=\frac{\Delta}{r^2+a^2+n^2}.
\end{equation}
As usual, the tortoise coordinate ranges from $-\infty $ to $+\infty$ when $r$ varies from the horizon $r_{\text{h}}$ to infinity, and thus covers the whole space outside the horizon.
Then, the radial equation becomes
\begin{equation}\label{rad}
     \frac{d^2R_{lm}}{dr_*^2}+\frac{2r\Delta}{(r^2+ a^2+n^2)^2}\frac{dR_{lm}}{dr_*}+\left[(\omega-\frac{ma}{r^2+a^2+n^2})^2 -\frac{\mu^2 r^2+\lambda_{lm}}{(r^2+a^2+n^2)^2}\Delta \right]R_{lm} =0.
\end{equation}
Since we are more concerned with waves incident into the black hole horizon, it is convenient to investigate the radial equation near the horizon. Near the horizon, it can be written as
\begin{equation}\label{radia}
   \frac{d^2R_{lm}}{dr_*^2}+ \left(\omega-m\frac{a}{r_\text{h}^2+a^2+n^2}\right)^2R_{lm}=0,
\end{equation}
which can be rewritten as
\begin{equation}\label{radiala}
   \frac{d^2R_{lm}}{dr_*^2}+\left(\omega-m\Omega_\text{h}\right)^2R_{lm}=0.
\end{equation}
It has the solution
\begin{equation}\label{radaialsol}
   R_{lm}(r)\thicksim \exp[\pm i(\omega-m\Omega_\text{h})r_*].
\end{equation}
The positive and minus signs correspond to the outgoing and ingoing wave modes, respectively. Requiring ingoing waves at the horizon, which is a physically acceptable solution, we choose the negative sign.  Thus, the field near the horizon is
\begin{equation}\label{sol}
   \Psi=\exp[-i(\omega-m\Omega_\text{h})r_*]S_{lm}(\theta)e^{im\phi}e^{-i\omega t}.
\end{equation}

Having the wave function, we can calculate the changes of the energy and angular momentum of the black hole through the flux of the scalar field.

\subsection{Conserved charges under scattering of the scalar field}

Since the physical origin of the NUT charge is still an open issue, we use the black hole thermodynamics to argue the parameter change of the black hole during the absorption of a test field or particle with energy $dE$ and angular momentum $dJ$.

We shoot a single wave mode $(l, m)$  into the Kerr-Taub-NUT black hole to investigate the changes of the parameters of the black hole. The energy and angular momentum carried by the scalar field can be estimated from their fluxes at the event horizon. The energy-momentum tensor corresponding to this wave mode is given by
\begin{equation}\label{energy-momentun}
   T_{\mu\nu}=\partial_{\left( \mu \right. } \Psi\partial_{\left. \nu \right)}\Psi^*
    -\frac{1}{2}g_{\mu\nu}\left(\partial_\alpha\Psi\partial^\alpha\Psi^* + \mu^2 \Psi \Psi^*  \right).
\end{equation}
The energy flux through the event horizon is \cite{Padm10}
\begin{equation}\label{energyflux}
   \frac{dE}{dt}=\int_H T^r_t\sqrt{-g} \, d\theta d\phi=\omega(\omega-m\Omega_\text{h})(r_\text{h}^2+a^2+n^2),
\end{equation}
and the angular momentum flux through the event horizon is \begin{equation}\label{ang}
   \frac{dJ}{dt}=-\int_H T^r_\phi\sqrt{-g} \, d\theta d\phi=m(\omega-m\Omega_\text{h})(r_\text{h}^2+a^2+n^2),
\end{equation}
where we have used the normalization condition of the angular functions $S_{lm}(\theta)$ in the integration \cite{Gwak18}. For waves with $\omega > m \Omega_\text{h}$, the energy and angular momentum flow into the event horizon; while, for waves with $\omega < m \Omega_\text{h}$, the energy and angular momentum fluxes are negative, which implies that waves with $\omega < m \Omega_\text{h}$ extract rotational energy from the black hole. This is called black hole super-radiance \cite{BrCP15}.

During an infinitesimal time interval $dt$, the changes in the mass and angular momentum of the black hole are
\begin{eqnarray}
   dM&=&dE = \omega(\omega-m\Omega_\text{h})(r_\text{h}^2+a^2+n^2) \, dt ,\\
   dJ &=&  m(\omega-m\Omega_\text{h})(r_\text{h}^2+a^2+n^2) \, dt.
\end{eqnarray}

For a black hole far from extremal, after the absorption of the infinitesimal energy and angular momentum of the  field, the final state is still a black hole. Then, we can use the black hole thermodynamics.

The Kerr-Taub-NUT black hole has three parameters, the mass $M$, angular momentum parameter $a$ and NUT parameter $n$.
If we assume the NUT parameter $n$ stays fixed in the absorption of a test field, from the first law of black hole thermodynamics,
\begin{equation}\label{firstlaw}
  dM=TdS+\Omega_\text{h} dJ+\psi dN,
\end{equation}
we have

\begin{eqnarray}
\nonumber  dA &=& \frac{4}{T}\left[(dM-\Omega_\text{h}dJ)-\psi dN\right] \\
\nonumber   &=& \frac{4}{T}\left[(\omega-m\Omega_\text{h})^2(r_\text{h}^2+a^2+n^2)dt -\frac{n^2}{2r_\text{h}^2}dr_\text{h}\right] \\
   &=& \frac{8\pi m^2r_\text{h}(r_\text{h}^2+a^2)}{2a^2n^2+ r_\text{h}(r_\text{h}-M)(r_\text{h}^2+a^2+n^2)}\left(\frac{\omega}{m}- \Omega'\right)\left(\frac{\omega}{m} -\Omega_\text{h}\right)(r_\text{h}^2+a^2+n^2)^2dt,
\end{eqnarray}
where we have defined
\begin{equation}\label{Omegap}
  \Omega'=\frac{2Mar_\text{h}}{(r_\text{h}^2+a^2+n^2)(r_\text{h}^2+a^2)}.
\end{equation}
It can be simplified as
\begin{equation}\label{Omegapp}
  \Omega'=\left(1-\frac{n^2}{r_\text{h}^2+a^2}\right)\Omega_\text{h}.
\end{equation}
It is clear that $\Omega'$ is smaller than the angular velocity of the event horizon. Hence, there exists an interval of wave modes
\begin{equation}\label{mode}
  \Omega'<\frac{\omega}{m}<\Omega_\text{h},
\end{equation}
such that the area of the event horizon decreases during the scattering process,
\begin{equation}\label{violat}
  dA<0.
\end{equation}
This leads to the violation of Hawking's area increasing theory, which states that during any classical process the area of a black hole event horizon never decreases \cite{Hawk72, CCFK18}. This suggests that the assumption for the fixed NUT parameter  in the  absorption of a test field with energy $\delta E$ and angular momentum $\delta J$ is inappropriate. Therefore, the NUT parameter should change during the scattering process.
This leads us to consider the thermodynamical charge $N$ is  fixed during the absorption process.

If we assume the thermodynamical charge $N$ is conserved during the scattering, from the first law of black hole thermodynamics, Hawking's area increasing theory is preserved naturally during the scattering process,
\begin{eqnarray}
\nonumber  dA &=& \frac{4}{T}(dM-\Omega_\text{h} dJ) \\
     &=& \frac{4}{T}(\omega-m\Omega_\text{h})^2(r_\text{h}^2+a^2+n^2)dt.
\end{eqnarray}
which is always positive. This indicates that the area of the event horizon increases during the scattering of the waves, and it is consistent with Hawking's area increasing theorem.

In the following discussion of weak cosmic censorship conjecture for the Kerr-Taub-NUT black hole, we will assume the thermodynamical charge $N$ does not change in the absorption of a scalar field or test particle with energy $dE$ and angular momentum $dJ$.

\section{Weak cosmic censorship conjecture for Kerr-Taub-NUT black hole with test scalar field}\label{4}

In this section, we examine the validity of the weak cosmic censorship conjecture by shooting a monotonic classical test scalar field with frequency $\omega$ and azimuthal harmonic index $m$ into the extremal and near-extremal Kerr-Taub-NUT black holes, and argue whether we can push the resulting composite object over the extremal limit, thus destroy the event horizon to form a naked singularity.

The event horizon of the black hole is determined by the metric function
\begin{eqnarray}
\Delta=r^2-2Mr+a^2-n^2,
\end{eqnarray}
with the minimum of $\Delta$
\begin{eqnarray}
\Delta_{\text{min}}=a^2-M^2-n^2 \label{Deltamin}
\end{eqnarray}
at the point $r_{\text{min}}=M.$  For a black hole, the minimum of the metric function $\Delta$ is negative or zero; while, for a naked singularity, it is positive and there is no solution for $\Delta=0$.

In the process of absorbing a test body (scalar field or particle) with energy $dE$ and angular momentum $dJ$, the changes of the parameters of the black hole are
\begin{eqnarray}
M &\rightarrow&  M'=M+dM,~~~~\nonumber\\
J &\rightarrow& J'=J+dJ, ~~~~\\
N &\rightarrow& N'=N.\nonumber
\end{eqnarray}
From the expressions (\ref{Deltamin}), (\ref{J}) and (\ref{N}), we have
%\begin{eqnarray}
%% \nonumber to remove numbering (before each equation)
%  \left(\frac{\partial\Delta_{\text{min}}}{\partial M}\right)_{J,N} &=& -\frac{4r_\text{h}(3a^2+n^2)+2M(3r_\text{h}^2+3a^2+n^2)}{3r_\text{h}^2+3a^2+n^2}, \\
%  \left(\frac{\partial\Delta_{\text{min}}}{\partial J}\right)_{M,N} &=& \frac{12ar_\text{h}}{3r_\text{h}^2+3a^2+n^2}.
%\end{eqnarray}
\begin{eqnarray}
  \left(\frac{\partial\Delta_{\text{min}}}{\partial M}\right)_{J,N} &=& -2\frac{\Upsilon}{\Theta}, \\
  \left(\frac{\partial\Delta_{\text{min}}}{\partial J}\right)_{M,N} &=& 12\frac{ar_\text{h}}{\Theta},
\end{eqnarray}
where
\begin{eqnarray}
   \Theta &=& {3r_\text{h}^2+3a^2+n^2},\\
   \Upsilon &=& 2r_\text{h}(3a^2+n^2)+M \Theta.
\end{eqnarray}
After the absorption of the test body, the minimum of the metric function $\Delta_{\text{min}}$ changes to $\Delta'_{\text{min}}$,
\begin{eqnarray}
\nonumber  \Delta'_{\text{min}} &=& \Delta'_{\text{min}}(M+dM, J+dJ, N) \\
\nonumber   &=& \Delta_{\text{min}} +\left(\frac{\partial\Delta_{\text{min}}}{\partial M}\right)_{J,N}dM+ \left(\frac{\partial\Delta_{\text{min}}}{\partial J}\right)_{M,N}dJ\\
   &=&  -(M^2+n^2-a^2)-\frac{2\Upsilon}{\Theta}dM +\frac{12ar_\text{h}}{\Theta}dJ.  \label{Delta'min1}
\end{eqnarray}

The value of the  event horizon is extremely close to the minimal point for a near-extremal black hole, and the value of  the event horizon coincides with the minimal point for an extremal black hole.

Now we consider the extremal and near-extremal black holes. The question, then, is whether $\Delta=0 $ has a positive solution after the black hole absorbs the test field, or equivalently, whether the minimum $\Delta_{\text{min}}$ of the metric function is positive.

Since the event horizon radius $r_{\text{h}}$ is extremely close to the minimal radius $r_{\text{min}}=M$ for the near-extremal black hole, we can define an infinitesimal distance $\epsilon$ between $r_{\text{h}}$ and $r_{\text{min}}$:
\begin{equation}
 r_\text{h}=r_{\text{min}}+\epsilon. \label{distance}
\end{equation}
We can see that $\epsilon >0$ and $\epsilon=0$ correspond to the near-extremal and extremal black holes, respectively. Before the absorption of the scalar field, the minimum of the metric function $\Delta$ can be written as
\begin{equation}
   \Delta_{\text{min}}=a^2-M^2-n^2=-\epsilon^2. \label{iminimal}
\end{equation}

Without loss of generality, we consider an infinitesimal time interval $dt$. For a long period of time, we can divide it into a lot of small time intervals $dt$, and consider the scattering process for each time interval separately by only changing the black hole parameters.

After the absorption of the scalar field, the minimum of the metric function $\Delta_{\text{min}}$ becomes $\Delta'_{\text{min}}$:
\begin{equation}\label{HorizonChecking}
     \Delta'_{\text{min}}= -(M^2+n^2-a^2)-\frac{2\Upsilon}{\Theta}dM +\frac{12ar_\text{h}}{\Theta}dJ.
\end{equation}
To first order in $dt$, we have
\begin{equation}
   \Delta'_{\text{min}}= -\epsilon^2-2m^2\frac{\Upsilon}{\Theta} \,
\left( \frac{\omega}{m}-\Omega \right) \left( \frac{\omega}{m}-\Omega_\text{h} \right) \, dt,  \label{Delta'min}
\end{equation}
where $\Omega_\text{h}$ is the angular velocity of the horizon defined in (\ref{angular}), and $\Omega$ is an effective angular velocity:
\begin{equation}
  \Omega=\frac{6ar_\text{h}}{\Upsilon}. \label{effangu}
\end{equation}
Now, we can check whether the Kerr-Taub-NUT black hole can be over-spun. This can be done by judging whether $\Delta'_{\text{min}}$ in (\ref{Delta'min}) is positive. For a naked singularity, $\Delta'_{\text{min}}>0$, while for a black hole, $\Delta'_{\text{min}}\leq 0$.

For the extremal Kerr-Taub-NUT black hole, the effective angular velocity and the angular velocity of the black hole are the same, i.e., $\Omega=\Omega_\text{h} $, and the minimum of $ \Delta' $ is
\begin{equation}\label{extrema}
  \Delta'_{\text{min}}=-\frac{24Mm^2 \, a^4 }{M^2+2a^2} \left( \frac{\omega}{m}-\Omega_\text{h} \right)^2 \,dt,
\end{equation}
which is always non-positive. For ${\omega}\neq {m}\Omega_\text{h}$, there will be two horizons after the absorption of the field and so the extremal Kerr-Taub-NUT black hole will become a non-extremal one after the scattering. While for ${\omega}={m}\Omega_\text{h}$, it will still be extremal. This indicates the extremal Kerr-Taub-NUT black hole cannot be over-spun.

For a near-extremal Kerr-Taub-NUT black hole, with the expressions (\ref{effangu}) and (\ref{distance}), we have
\begin{equation}\label{Omega2}
  \Omega=\Omega_\text{h}+\frac{4(3M^2+n^2)\epsilon +18M\epsilon^2+6\epsilon^3}{\Upsilon},
\end{equation}
which shows that the effective angular velocity $\Omega$ is larger than the angular velocity of the black hole, $\Omega>\Omega_\text{h}$.
%\begin{equation}
%      \left( \frac{\omega}{m}-\Omega_\text{h}\right) \left( \frac{\omega}{m} -\Omega \right)+\frac{\left(\Theta\right)\epsilon^2}{2m^2 \,\left[2r_\text{h}(3a^2+n^2)+M(3r_\text{h}^2+ 3a^2+n^2)\right](r_\text{h}^2+a^2+n^2) \, dt}<0. \label{aa}
%\end{equation}
It is clear that, for wave modes with
\begin{eqnarray}
  \frac{\omega}{m}=\frac{\Omega_\text{h}+\Omega}{2}, \label{Omega/m}
\end{eqnarray}
the value of $\Delta'_{\text{min}} $ is the largest. Thus, if these wave modes can not over-spin the near-extremal Kerr-Taub-NUT black hole, all the wave modes can not over-spin the near-extremal black hole either.
We shoot one of these wave modes into the near-extremal black hole. Then, by substituting (\ref{Omega2}) and (\ref{Omega/m}) into (\ref{Delta'min}), we have
\begin{equation}
   \Delta'_{\text{min}}= -\epsilon^2
    +
    \frac{8m^2(3M^2+n^2)^2  }
         {{\Theta}\,
          \Upsilon}
    \left[\epsilon^2 + \mathcal{O}({\epsilon^3})\right]\, dt,  \label{Delta'min2}
\end{equation}
By choosing the infinitesimal time interval $dt <\epsilon$, we can see that
\begin{equation}
   \Delta'_{\text{min}} < -\epsilon^2
    +
    \frac{8m^2(3M^2+n^2)^2  }
         {{\Theta}\,
          \Upsilon}
    \left[\epsilon^3 + \mathcal{O}({\epsilon^4})\right] <0,
\end{equation}
which shows that it is impossible to form a naked singularity and the event horizon cannot be destroyed.

Thus, both the extremal and near-extremal Kerr-Taub-NUT black holes cannot be over-spun by test scalar fields. Hence, the weak cosmic censorship conjecture is preserved.

\section{Weak cosmic censorship conjecture for Kerr-Taub-NUT black hole with test particle}\label{5}

Another method  to check weak cosmic censorship conjecture is throwing a test particle with large angular momentum into the extremal or near-extremal black hole. This gedanken experiment was first proposed by Wald~\cite{Wald74}, and further developed by Hubeny, Jacobson and Sotiriou~\cite{Hube99,JaSo09}. It was shown that the event horizon of a near-extremal Reissner-Nordstr\"om black hole or a Kerr black hole can be destroyed \cite{Hube99,JaSo09}. In this section, we use this method to check whether the event horizon of a Kerr-Taub-NUT black hole can be destroyed.

A test particle with rest mass $m$ moving in the Kerr-Taub-NUT space-time can be described by the geodesic equation
\begin{equation}
  \frac{d^2x^\mu}{d\tau^2}+ \Gamma^\mu_{\alpha\beta}\frac{dx^\alpha}{d\tau}\frac{dx^\beta}{d\tau}=0,
\end{equation}
which can be derived from the Lagrangian
\begin{equation}\label{Lagr}
  L= \frac{1}{2}mg_{\mu\nu}\frac{dx^\mu}{d\tau}\frac{dx^\nu}{d\tau}.
\end{equation}
The energy $\delta E$ and angular momentum $\delta J$ of the particle are
\begin{eqnarray}
  \delta E &=& -P_t=-\frac{\partial L}{\partial \dot{t}}= -mg_{0\nu}^{~}\frac{dx^\nu}{d\tau}, \label{deltaEP} \\
  \delta J &=& P_\phi=\frac{\partial L}{\partial \dot{\phi}}=mg_{3\nu}^{~}\frac{dx^\nu}{d\tau}. \label{deltaJP}
\end{eqnarray}

We first find the condition for a particle with energy $\delta E$ and angular momentum $\delta J$ to enter the black hole, and then check whether such particle violates the weak cosmic censorship conjecture.

The four velocity of a massive particle is a time-like and unit vector,
\begin{equation}\label{fourV}
  g_{\mu\nu}\frac{dx^\mu}{d\tau}\frac{dx^\nu}{d\tau}=\frac{1}{m^2}g^{\mu\nu}P_\mu P_\nu =-1.
\end{equation}
Substituting the energy $\delta E$ (\ref{deltaEP}) and angular momentum $\delta J$ (\ref{deltaJP}) into the above equation, we get
\begin{equation}\label{EJM}
  g^{00}\delta E^2-2g^{03}\delta J\delta E+g^{11}P_r^2+g^{22}P_\theta^2+g^{33}\delta J^2 = -m^2.
\end{equation}
Then the energy of the particle is
\begin{equation}\label{EnergyP}
  \delta E=\frac{g^{03}}{g^{00}}\delta J- \frac{1}{g^{00}}\left[(g^{03})^2\delta J^2-g^{00}g^{33}\delta J^2- g^{00}(g^{11}P_r^2+g^{22}P_\theta^2+m^2)\right]^{\frac{1}{2}}.
\end{equation}
Since the motion of a massive particle outside the event horizon should be future directed and time-like, we have chosen the future directed solution $dt/d\tau>0$, which is equivalent to the requirement
\begin{equation}\label{Condition}
  \delta E>-\frac{g_{03}^{~}}{g_{33}^{~}}\delta J.
\end{equation}
If the particle enters the black hole, it must cross the event horizon. On the event horizon, the condition becomes
\begin{equation}\label{horicond}
  \delta E>\frac{a}{r_\text{h}^2+a^2+n^2}\delta J =\Omega_\text{h} \delta J.
\end{equation}
Thus, for the particle to be absorbed by the black hole, the angular momentum  of the particle must satisfy
\begin{equation}
  \delta J<\delta J_{\text{max}}=\frac{1}{\Omega_\text{h}}\delta E. \label{ConditoinJ1}
\end{equation}
On the other hand, in order to over-spin the black hole, the minimum of the metric function $\Delta$ should be positive after the absorption of the particle. To first order, the condition is
\begin{equation}\label{min}
  \Delta'_{\text{min}}= -(M^2+n^2-a^2)-\frac{2\Upsilon}{\Theta}\delta M +\frac{12ar_\text{h}}{\Theta}\delta J>0,
\end{equation}
which is come from (\ref{Delta'min1}) and can be rewritten as
\begin{equation}
  \delta J>\delta J_{\text{min}}=\frac{1}{\Omega}\delta E +\frac{\Theta}{12ar_\text{h}}(M^2+n^2-a^2). \label{ConditoinJ2}
\end{equation}
When the two conditions (\ref{ConditoinJ1}) and (\ref{ConditoinJ2}) are satisfied simultaneously, the black hole can be over-spun and weak  cosmic censorship conjecture is violated.
%the question is whether there exists particle with energy $\delta E$ such that $\delta J_{\text{max}}>\delta J_{\text{min}}$.

For the extremal Kerr-Taub-NUT black hole, we have $M^2+n^2-a^2=0$ and $\Omega=\Omega_\text{h}$.
Therefore,
 \begin{eqnarray}
   \delta J_{\text{max}} = \frac{1}{\Omega_\text{h}}\delta E =\delta J_{\text{min}},
 \end{eqnarray}
which means that the two conditions (\ref{ConditoinJ1}) and (\ref{ConditoinJ2}) can not be satisfied simultaneously. Thus, the weak cosmic censorship conjecture is preserved for the extremal Kerr-Taub-NUT black hole.

For the near-extremal Kerr-Taub-NUT black hole, as indicated by Eq. (\ref{Omega2}), the effective angular velocity is larger than the angular velocity of the black hole, i.e., $\Omega > \Omega_{\text{h}}$. To first order, obviously there exist particles with energy $\delta E$ such that $\delta J_{\text{max}}>\delta J_{\text{min}}$, which shows that the two conditions (\ref{ConditoinJ1}) and (\ref{ConditoinJ2}) can be satisfied simultaneously.
%Fig.(\ref{fig2}) shows an example violating the weak cosmic censorship conjecture.
Thus, the near-extremal Kerr-Taub-NUT black hole can be over-spun by the test particle.

%Hence, to first order, the extremal Kerr-Taub-NUT black hole cannot be over-spun by the test particle; while the near-extremal Kerr-Taub-NUT black hole can be over-spun.

\section{Discussion and Conclusions}\label{6}

In this paper, we have investigated the validity of the weak cosmic censorship conjecture for a Kerr-Taub-NUT black hole by test scalar fields and particles. For the test scalar field scattering gedanken experiment, we considered an infinitesimal time interval. The result suggests that both extremal and near-extremal Kerr-Taub-NUT black holes cannot be over-spun. For the test particle thought experiment, the study suggests that extremal Kerr-Taub-NUT black hole cannot be over-spun; while near-extremal Kerr-Taub-NUT black hole can be over-spun.
Although the first-order approximation was considered in our procedure, it has strong evidence that, for the test particle approximation, the result might still be the same as in the Kerr black hole case \cite{JaSo09}.

%In the scalar field scattering process, we considered the time that field enters into the black hole naturally through the calculation of energy and angular momentum flux. While,  in the test  particle gedanken experiment, the  time interval that the particle cross the horizon is instant. For consideration of delicate problems like the weak cosmic censorship conjecture, the time that particle crosses the horizon might be important as indicated by Gwak \cite{Gwak17,Gwak18,LiWL19}.

\section*{Acknowledgements}

 This work was supported in part by the National Natural Science Foundation of China (Grants No. 11875175, No. 11522541, No. 11675064, and No. 11875151), and the Fundamental Research Funds for the Central Universities (Grants No. lzujbky-2018-k11 and No. lzujbky-2017-it68).

\end{document}